# Optimized Absorption of Light in Perovskite Nanowire Solar Cells

Sina Salimian[1] and Nicklas Anttu[1,*]

[1]Physics, Faculty of Science and Engineering, Åbo Akademi University, FI-20500 Turku, Finland

*E-mail: nicklas.anttu@abo.fi

**Abstract**

Metal halide perovskite nanowires (PrvNWs) have recently emerged as an interesting path for nanostructured solar cells. Here, we model the absorption of light in PrvNW arrays for varying diameter and length of the PrvNWs and period for the array by solving the Maxwell equations. For long enough bare PrvNW arrays, we find that the optimum diameter is fixed to that which places the absorption peak from the $HE_{11}$ waveguide mode in the PrvNWs to the vicinity of the bandgap wavelength. In contrast, when we include a transparent conductive oxide (TCO) top contact layer, the optimum diameter shifts to a larger value by 100 nm. The origin of this shift is traced to a reduced reflection at the interface between the TCO layer and the PrvNW array when the PrvNW's diameter is larger. Overall, we find that 1500 nm long PrvNWs can reach 90% of the broadband absorption potential, making this system of high interest for photovoltaics.

**Introduction**

Solar cells, first introduced in 1954 in the context of a practical device [1], have been demonstrated as the defining paradigm of renewable energy harvesters, attracting significant attention during the past couple of decades. The increasing ease in mass-production of electronic devices, in combination with fossil-fuel induced $CO_2$-driven global warming that threatens humankind's future, makes solar cell technology a highly demanded one to be investigated both academically and industrially [2]. Technically, the working principle of solar cells is based on the conversion of solar energy into electricity, which is accomplished by the absorption of light on a photoactive cell, giving rise to the generation of excess electron-hole pairs, and then, the separation of those charge carriers into an external circuit at a voltage bias [3]. The characteristics of these cells can be viewed from several directions such as engineering (fabrication process), chemistry (material selection and characterization), and physical (conversion of light into free charges) [2]. Currently, more than 4.5% of the world's global electricity is produced via solar cells [4], which makes optimizing their performance, i.e., power conversion efficiency (PCE), more relevant than ever.

From the engineering aspect, the conventional, market-dominating solar cells (known as single-junction solar cells) are configured as a single layer of light-absorbing semiconductor material (known as the active layer) [5]. The active layer is sandwiched between two additional layers playing the role of cathode and anode interlayers for facilitating hole and electron collection [6]. Currently, the lab efficiency is over 26% for the industry-dominating silicon active layer [7], which can be enhanced to over 29% in the case of GaAs active layer [8]. However, the development pace of single junction semiconductor solar cells has become rather slow in recent years due to the limited spectral response available from the single active layer and fabrication costs if aiming to increase the electrical performance of the device [9, 10]. A potential solution to overcome the limited PCE is the use of tandem solar cells (TSCs), which consist of multiple active layers stacked on top of each other, allowing a great enhancement in efficiency and broader spectrum absorption due to the different bandgaps of each layer [11]. By employing such a configuration, the reported PCE is drastically enhanced, increasing to 35.5% for two-junction [12], and even to the point of 47.6% for four-junction solar cells [8]. The challenge here is the more complicated fabrication process and typically much higher cost of fabrication with increasing number of active layers. A tempting research direction is then to enhance the efficiency of TSCs instead of increasing the number of junctions, which motivates the direction of the present work.

From the chemistry aspect, metal halide perovskites and their integration in solar cells, better known as perovskite solar cells (PSCs), have made a huge and very swift progress during the past decade, from its first solid-state emerging solar cell introduced in 2012 with PCE already over 10% [13] to the latest lab efficiency report of 26.0% for a single junction PSC and 33.7 % for a perovskite/silicon TSC [14]. As organic/inorganic hybrids, perovskites have gained a key position in the solar cells industry due to their tunable optical properties, efficient light absorption coefficient, long carrier diffusion length, and low-cost fabrication methods rooted in their solution processability, which is pushing PSCs' large-scale production forward [15, 16]. These high-performance optoelectronic properties, in addition to their promising eco-friendliness and their integration flexibility, make PSCs excellent candidates as a major future direction in world-wide photovoltaics [17].

From the physical aspect, employing perovskite nanomaterial shaped into nanowires (PrvNWs) has been described as a highly promising direction in optoelectronic devices [18]. Some studies addressed their non-toxic integration in photodetectors [19] and strategies for stabilizing their crystal phase [20, 21] with promising results, and also the prospect of passivating the surface for reducing non-radiative recombination [22, 23]. Indeed, the possibilities of reducing the defect density and high carrier diffusion lengths in PrvNWs make them an interesting alternative to conventional perovskite thin-film devices [24]. Additionally, PrvNWs enable flexible thin-film photovoltaics [25]. Importantly, previous research has shown strong tunability of III-V and silicon NWs' optical response to enhance absorption [26]. We therefore expect that an optimized 3D geometry of the PrvNWs could offer optical benefits as compared to the optical response in conventional perovskite thin-films [27]. Inspired by bottom-up fabricated high-efficiency single-junction III-V NW array solar cells [28], we can aim to design strongly absorbing PrvNW-based PSCs that will be both cost-effective and also exceptionally lightweight to be employed as wearable renewable energy harvesters.

In this paper, we theoretically studied the optical properties of an array of PrvNWs and their dependence on the geometrical parameters of the PrvNWs and the array. For a constant length for the NWs, we determined the optimum diameter and period for absorption of sunlight, in terms of the short-circuit current ($J_{sc}$). The $J_{sc}$ is obtained from the modelled absorption spectrum in the PrvNWs.

With increasing NW length, the optimum period increases, whereas the diameter stays at a rather constant value. For the systems we consider before the inclusion of electrode layers, the diameter-independence originates from the positioning of a diameter-dependent absorption peak near the bandgap wavelength. We found that an increase of the refractive index of the material between the PrvNW's can have a beneficial effect on the absorption and, subsequently, $J_{sc}$ especially for shorter PrvNWs. Thus, since the practical PrvNW length available for devices typically depends heavily on the quality of the fabrication process, the choice of the material planarizing the PrvNW array should be considered in detail in the design process.

Lastly, for a more practical configuration for solar cells, we investigated PrvNWs sandwiched between two transparent-conductive oxide (TCO) electrodes. We found that for our choice of the TCO, the absorption in the PrvNWs and optimum $J_{sc}$ are noticeably limited by the reflection and the absorption of the top electrode, which remain rather constant with increasing length. Thus, we recommend a careful consideration of including an anti-reflection layer and/or changing the thickness or material of the electrode to unlock the full potential of NW-based PSCs. Furthermore, we found for this case a considerably higher optimum diameter than for the cases without the presence of TCO layers; the origin for this larger diameter is rooted in a better refractive index matching between the TCO layer and the PrvNW layer at a larger NW diameter.

## Methods

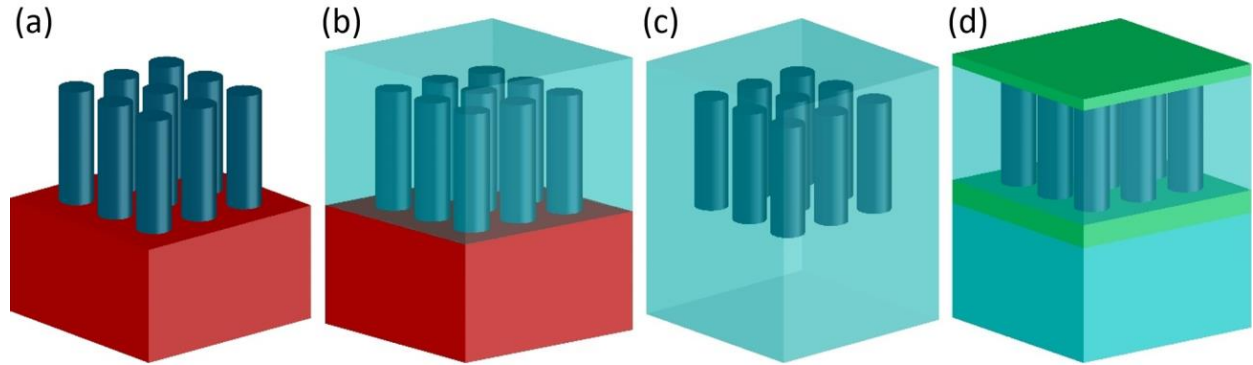

**Figure 1.** Schematic of PrvNWs in the four systems that we consider, that is, (a) on a silicon substrate surrounded by air, (b) on a Si substrate covered by a surrounding material with refractive index of 1.5, (c) as (b) but here also the substrate has a refractive index of 1.5, and (d) sandwiched between two electrodes on a substrate with refractive index of 1.5.

We chose to model the perovskite $CH_3NH_3PbI_3$, with the refractive index extracted from Ref. [29] (see figure S1 in the Supplementary Information for the refractive index values used). $CH_3NH_3PbI_3$ is currently the best-performing perovskite for solar cells [30], with a bandgap corresponding to approximately 789.9 nm (1.57 eV) [29]. We model the light scattering in a 3-D period array of

PrvNWs in four different systems. In all systems, the nanowires are of diameter $D$ and length $L$; and placed in a square array of period $P$. First, we consider a design where the PrvNWs are placed on top of a silicon substrate and otherwise surrounded by air ($n = n_s = 1$) (figure 1(a)). Second, we change the surrounding medium from air to one with higher refractive index ($n = n_s = 1.5$, typical for polymers or oxides used for planarizing nanowire arrays), still on a Si substrate (figure 1(b)). Third, we change also the substrate to $n = n_s = 1.5$ (figure 1(c)). As our last, fourth, system (figure 1(d)), to analyze the PrvNWs in an even more realistic context for PSCs, we sandwiched the PrvNW array between two layers of fluorine-doped tin oxide (FTO) [31] which is widely known to be an excellent choice for the fabrication of the transparent conductive electrodes in PSCs (see figure S2 in the Supplementary Information for the refractive index values used). In this system, the material above the top FTO layer had $n = 1$ and the material below the bottom FTO layer had $n = 1.5$ to resemble a mechanically supporting membrane for the otherwise very thin film.

The scattering of light is modelled with the Maxwell equations [32], and we use Lumerical FDTD Solutions for the numerical solving with the finite-difference time-domain (FDTD) method. In the simulations, an incident plane wave source is radiating at a normal angle to the PrvNW array with the polarization direction along the $x$-axis, covering the wavelength range from 270 to 790 nm. The simulation time is set at a maximum of 300 fs, with automatic early shut-off if we reach a relative $10^{-5}$ level for the power remaining within the simulation domain. The $z$-axis boundary conditions are set to perfectly match layers (PMLs). For the first and second (third and fourth) systems, we simulate a thickness of 1 μm of the silicon substrate (a substrate with refractive index of 1.5) substrate, with the PML continuing into the substrate, effectively corresponding to an infinitely thick substrate. We use symmetric boundary conditions for periodic structures along the $x$- and $y$-axis (asymmetrical along the $x$-axis, symmetrical along the $y$-axis, on both sides for each axis) to ensure the periodicity of the arrangement while reducing the volume of the simulation domain by a factor of four, and hence also the computation time needed by approximately a factor of four. We use a uniform mesh with a 10 nm step between grid points in each spatial direction.

For the first three systems (see figure 1(a)-(c) for schematics), the absorption in the PrvNWs is extracted according to $A = 1 – (R + T)$ from two different 2-D monitors perpendicular to the z-axis, placed above and below the PrvNW array, which are used for calculating the reflection ($R$) and transmission ($T$), respectively (note that this $T$ is the transmission into the substrate that continues toward infinity in the downward direction). For the fourth system (figure 1(d)), four monitors were considered, at the top and the bottom interfaces of each FTO layer, to enable separate assessment of the absorption in the PrvNWs, in the top FTO layer, and the bottom FTO layer, as well as the reflection loss and the loss due to transmission into the substrate.

To quantify the broadband absorption in a manner relevant for solar cells, we calculate the short-circuit current, under the assumption that each absorbed above-bandgap photon in the PrvNWs gives rise to one charge carrier in the photocurrent, similarly as in [33] for III-V NWs:

$$J_{sc} = q \int_{\lambda_{short}}^{\lambda_{bg}} \frac{I_{AM1.5}(\lambda)A(\lambda)}{2\pi\hbar c/\lambda} d\lambda. \tag{1}$$

Here, $A$ is the absorption spectrum of the PrvNWs, $q$ the elementary charge, $\hbar$ the reduced Plank constant, $c$ the speed of light in vacuum, $I_{AM1.5}$ the AM1.5G incident solar spectrum [34], $\lambda_{bg}$ is the bandgap of the PrvNWs (789.9 nm), and $\lambda_{short} = 270$ nm is a lower limit below which the AM1.5G spectrum shows negligible intensity. Note that we can in Eq. (1) instead of $A(\lambda)$ use $R(\lambda)$ or $T(\lambda)$ to assess how much of the $J_{sc}$ potential is lost due to reflection or transmission; and similarly, by using the absorption spectrum of the FTO layers for the fourth system, we can assess how much of the incident light is lost due to absorption in the top and bottom FTO layer, respectively.

In our study, for all the four systems in figure 1, we varied $D$ with a 10 nm step within the range of 10 nm $\leq D \leq P$, and $P$ was varied in step of 10 nm up to 1000 nm. For PrvNW length, we considered $L = 500, 1000, 1500, 2000, 3000, 4000$, and 5000 nm. In this way, we can find for each of the four systems the optimum $D$ and $P$, denoted as $D_O$ and $P_O$, as a function of $L$.

## Results

We start by considering an array of PrvNWs with length of 500 nm, surrounded by air, and placed on a silicon substrate (see figure 1(a) for a schematic). The optimum absorption spectrum for this length is shown in figure 2 which occurs for $D_O = 260$ nm $P_O = 360$ nm. The absorption maintains a high level of $> 0.8$ until a sudden drop in the close vicinity of $\lambda_{bg} \approx 790$ nm where the absorption strength of the material drops toward zero (see figure S1 in the Supplementary Information).

We find that the PrvNWs absorb poorly when the diameter is small (figure 3(a) where $P = P_0$ is kept constant). This weak absorption is not merely due to less absorbing material in the array, but also due to the electrostatic screening of normally incident light in small-diameter nanowires [35]. With increasing $D$, the absorption increases. For a fixed diameter in the range $130 < D < 250$ nm, we find a distinct red-shifting $\lambda$ at which the absorption shows a peak (eventually, the absorption experiences a sudden drop when approaching close enough to $\lambda_{bg}$). This peak in the absorption is due to efficient coupling of light into and subsequent strong absorption through the fundamental $HE_{11}$ waveguide mode of the PrvNWs, in line with previous studies on III-V nanowires [26]. At $D \approx 250$ nm, the absorption peak due to the $HE_{11}$ mode is in the vicinity of the bandgap, giving rise to the strong overall absorption for $\lambda < \lambda_{bg}$ found in figure 2 when $P$ is optimized. Lastly, figure 3(b) shows, for $L = 500$ nm, the short circuit current as a function of diameter and period, which results in the optimum $J_{sc}$ of 21.5 mA/cm$^2$ for the $D_O = 260$ nm $P_O = 360$ nm considered in figure 2.

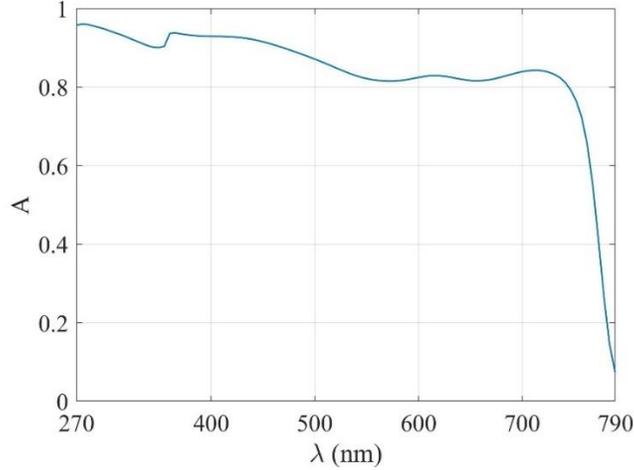

**Figure 2.** The absorption spectrum of PrvNWs surrounded by air, with $D_O = 260$ nm and $P_O = 360$ nm, which are the optimized values for the $L = 500$ nm considered here.

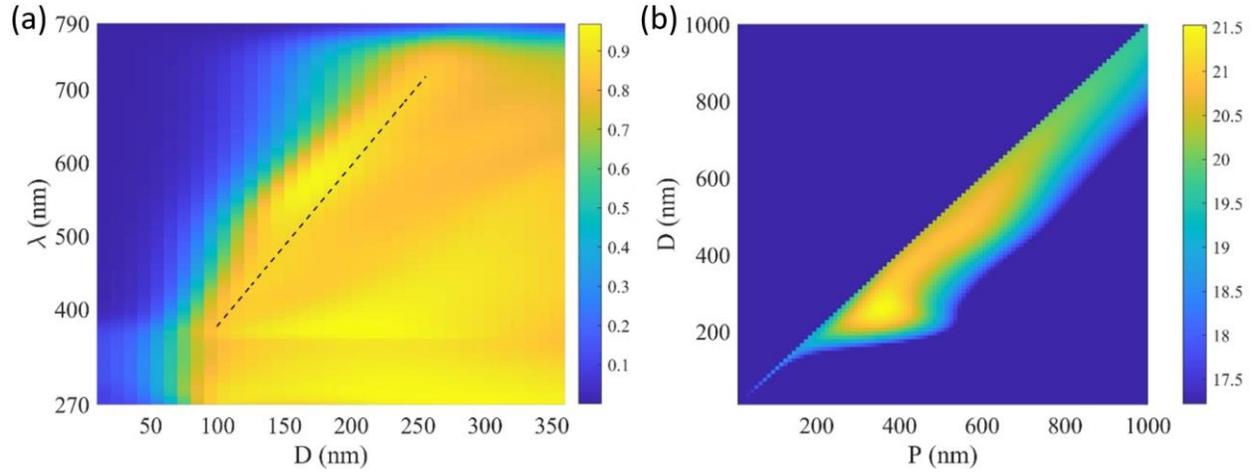

**Figure 3.** (a) The absorption in PrvNWs in the first system (figure 1(a)) as a function of diameter and wavelength at $P = P_O = 360$ nm and $L = 500$ nm. The straight diagonal line (dashed black line) indicates the strong absorption caused by the $HE_{11}$ waveguide mode, with weak absorption caused by electrostatic screening for small-diameter NWs. (b) $J_{sc}$ in mA/cm$^2$ as a function of $D$ and $P$ at $L = 500$ nm, with the absorption spectrum for the optimum dimensions shown in figure 2.

Next, for this first system (figure 1(a)), we extracted the optimum $P_O$ and $D_O$, with corresponding $A$ and $J_{sc}$, as a function of $L$ (for the $D$ and $P$ dependence of $J_{sc}$ for each $L$ considered, see figure S3 in the Supplementary Information). As $L$ increases, the optimum $A$ saturates at all wavelength toward 100% (figure 4), as expected from the increasing amount of absorbing perovskite material available, under the assumption that insertion reflection loss does not limit absorption with increasing $L$. Indeed, we find that $P_O$ increases with $L$ (figure 5(a)), which leads to a decrease in reflection loss, since $D_O$ stays rather constant as seen in figure 5(a) and the insertion reflection loss tends to decreases with decreasing $D/P$ ratio (see figure S7 in the Supplementary Information). The rather constant $D_O$ with increasing $L$ indicates that the optimum diameter is set by placing the $HE_{11}$ absorption peak to the vicinity of the bandgap, while the increasing $L$ allows to increase $P$, which in turn decreases the insertion reflection loss [26]. The increase in optimum $J_{sc}$ with increasing $L$ is shown in figure 5(b), where we also show the maximum possible $J_{scmax}$ (26.5 mA/cm$^2$, as obtained from Eq. (2) when using $A(\lambda) = 1$ for $\lambda \leq \lambda_{bg}$). As can be seen, already for $L$ = 1500 nm, we have reached 90% of $J_{scmax}$, and with increasing $L$, we converge toward $J_{scmax}$. Thus, these PrvNWs appear rather promising for absorption applications.

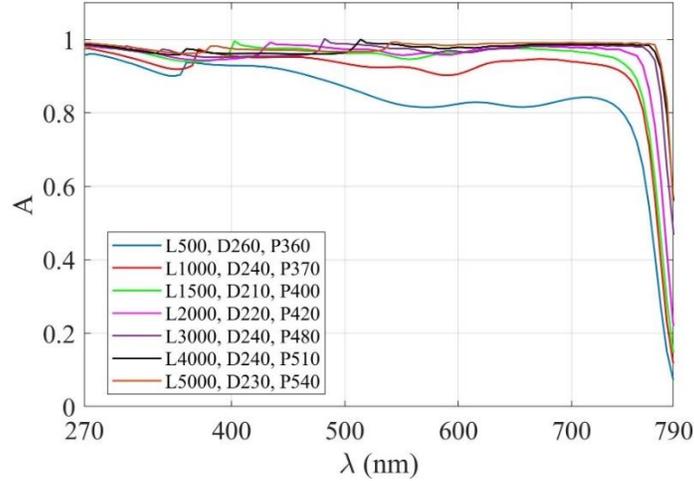

**Figure 4.** Absorption for varying $L$ at the $D$ and $P$ that optimize the absorption, in terms of $J_{sc}$, at that given $L$. That is, we show results for $D = D_O$ and $P = P_O$ here. The small and sharp peaks where $A \rightarrow 1$ are due to the Rayleigh anomalies where a diffracted order in the air at the top of the PrvNW array propagates parallel to the interface [36].

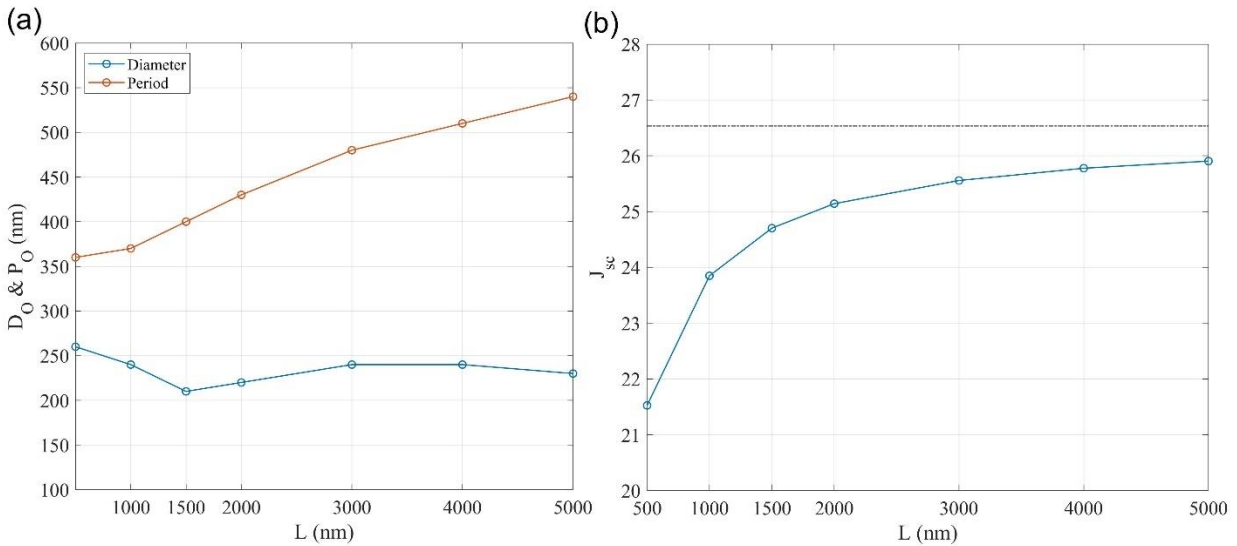

**Figure 5.** (a) $D_O$ and $P_O$ as a function of the length $L$ of PrvNWs in the first system (figure 1(a)). (b) The dependence on $L$ for the optimum $J_{sc}$, calculated with Eq. (1) from the $A$ in figure 4. The dashed-dotted line shows $J_{scmax} = 26.5$ mA/cm$^2$.

Next, we changed the air that covered the PrvNWs above to a surrounding with $n = n_s = 1.5$ and we consider two cases for the substrate; first with a silicon substrate (see figure 1(b)) and then with $n_s = 1.5$ also for the substrate (see figure 1(c)). Our purpose was to assess the impact of reducing the refractive index contrast between PrvNWs and their surroundings, compared to the case of $n = 1$ above, as would occur if planarizing the array with an oxide or a polymer.

The optimum $J_{sc}$ of the PrvNW array of 500 nm length on silicon substrate, acquired at $D_O = 330$ nm and $P_O = 390$ nm (see figure S4 in the Supplementary Information for $D$ and $P$ dependence), is 22.2 mA/cm$^2$, which is more than 0.7 mA/cm$^2$ higher than the above considered original geometry with nanowires surrounded by air above and between. Figure 6(a) shows for $P = P_O$ a rather similar diameter dependence as in the original setup (figure 1(a)), with low absorption level in the 50-150 nm diameter range and absorption edge due to the HE$_{11}$ mode. When we turn to the case with $n = 1.5$ also under the PrvNWs, $J_{sc}$ (extracted at $D_O = 380$ nm and $P_O = 440$ nm, see figure S5 in the Supplementary Information for $D$ and $P$ dependence) shows a noticeable increase of more than 1.8 mA/cm$^2$ in comparison to the original setup. Additionally, the absorption in figure 6(b) shows much richer features compared to the cases with silicon substrate. We assign these richer features to excitation of in-plane guided optical modes in the nanowire matrix [37]. The dispersion of such modes is strongly dependent on $P$, in addition to $D$, since they are partly localized to the space between the nanowires [37]. The reason why absorption features from such modes are not seen in the presence of the silicon substrate is that the modes can efficiently leak into the high refractive-index substrate; with the substrate absent in figure 6(b), strong excitation of the in-plane guided modes becomes possible since the modes remain much more bound to the array [37].

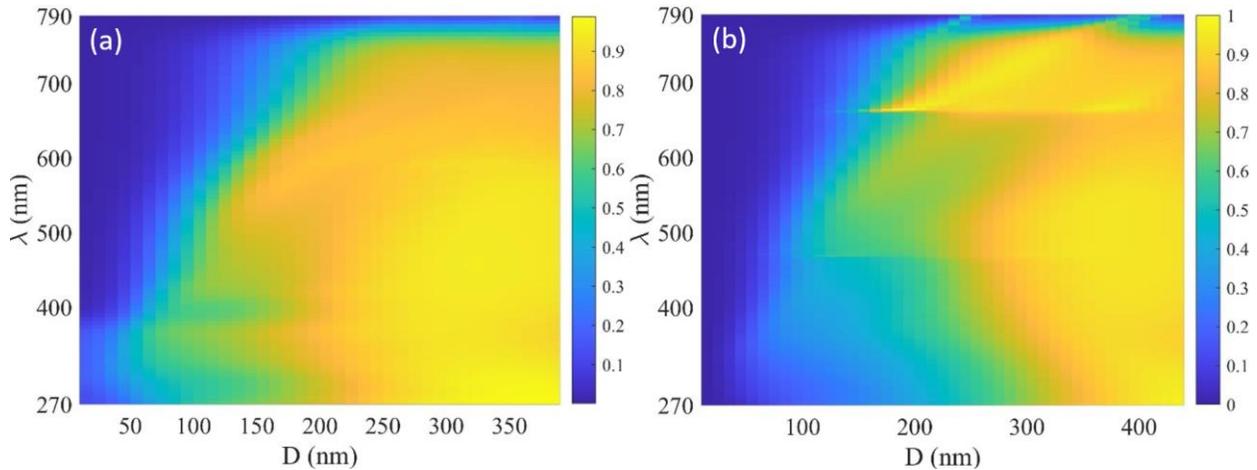

**Figure 6.** The absorption as a function of wavelength and diameter of the PrvNWs. (a) With a surrounding of $n = 1.5$ and a silicon substrate—see figure 1(c) for schematic. (b) Like (a) but with a substrate of also $n = 1.5$.

The comparison of the optimum absorption in the three systems above is shown in figure 7. For long enough nanowires, the difference between the systems is small; for example, at $L = 5000$ nm, the difference in optimum $J_{sc}$ is less than 0.1 mA/cm$^2$ between the systems, and the optimum $D$ and $P$ are rather similar for all three cases. However, for shorter nanowires, we find considerably higher $J_{sc}$ when increasing the $n$ of the material between and above the nanowires. Thus, since there typically is a limit on the length of NWs in practical applications due to the NW quality obtainable from the fabrication process, the refractive index of the surrounding material can be used as an additional design parameter to allow for a smaller $L$ to achieve a certain level of absorption and $J_{sc}$.

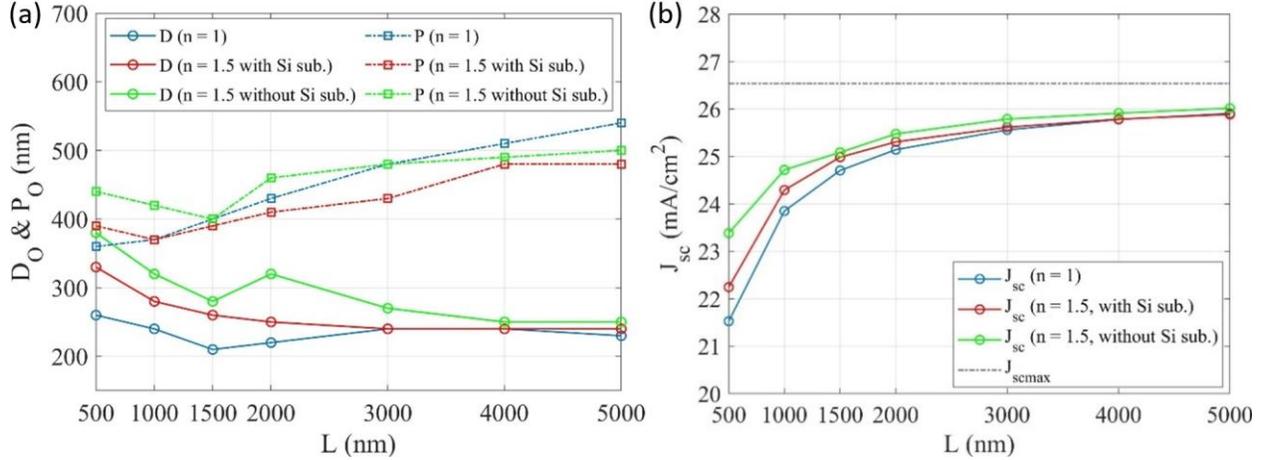

**Figure 7.** (a) $D_O$ and $P_O$ for PrvNWs in the three systems shown in figure 1(a)-(c). (b) The corresponding $J_{sc}$, for each length (see figure S8 in the Supporting Information for details of reflection and transmission loss in $J_{sc}$).

As the final step, we considered the case where FTO layers, intended as electrodes in practical solar cells, are included (see figure 1(d) for schematic). The top FTO layer's thickness was set to 100 nm while 200 nm was used for the bottom layer. This difference in layer thicknesses was motivated by the fact that for the top layer, we wish high transparency, even if at a slight cost of increased electrical resistance, while for the bottom contact, we can allow a larger thickness to decrease resistive losses. A substrate with $n_s = 1.5$ was placed below the bottom FTO layer as the supporting membrane (for the real and imaginary part of the refractive index of the FTO used in the simulations, see figure S2 in the Supplementary Information)

For the analysis, in addition to $D_O$, $P_O$, $A$, and $J_{sc}$ of the PrvNWs for varying $L$, we analyzed also the behavior of the FTO layers and reflection/transmission losses for the optimum geometry (for a schematic of this analysis, see figure S9 in the Supplementary Information). For the loss analysis, we have a general balance in short-circuit current according to:

$$J_{scmax} = J_{sc,R} + J_{sc,FTOtop} + J_{sc} + J_{sc,FTObottom} + J_{sc,T} \tag{2}$$

where, with Eq. (1), $J_{sc,R}$ is calculated from the reflection at the interface of air and the top FTO layer, $J_{sc,FTOtop}$ from the absorption in the top FTO layer, $J_{sc,FTObottom}$ from the absorption in the bottom FTO layer, and $J_{sc,T}$ from the transmission at the interface of the bottom FTO and substrate (and $J_{sc}$ is also here calculated from the absorption in the PrvNWs).

The results of this short-circuit analysis is shown in figure 8(a) (see figure S6 in the Supporting Information for the $D$ and $P$ dependence for each $L$ considered). For $L$ = 500 nm, $J_{sc,R}$ and $J_{sc,FTOtop}$ are 3.95 mA/cm$^2$ and 3.32 mA/cm$^2$, respectively, which remain rather constant with increasing $L$, as expected because of their planar geometry and the fact that light reaches those layers first (see figure S9 in the Supplementary Information for a schematic of the scattering of light inside the system). The $J_{sc}$ is 17.3 mA/cm$^2$ for $L$ = 500 nm and reaches 19.4 mA/cm$^2$ for $L$ = 5000 nm.

For $L$ = 500 nm, $J_{sc,FTObottom}$ = 0.85 mA/cm$^2$ and $J_{sc,T}$ = 1.13 mA/cm$^2$. With increasing $L$, both these terms reduce in magnitude, and at $L$ = 5000 nm, they are below 0.03 mA/cm$^2$. This decrease is expected since with increasing $L$, less light reaches the bottom FTO layer for absorption or transmission (see figure S9 in the Supplementary Information for schematic of light scattering in the system).

Importantly, we thus find that the $J_{sc}$ is limited by approximately 4 mA/cm$^2$ due to the reflection from the top FTO layer and 3 mA/cm$^2$ from absorption in the top FTO layer. These losses cannot be circumvented by making the nanowires longer. Therefore, to reach a higher $J_{sc}$, we recommend to use an anti-reflection coating on top of the FTO layer and a more transparent FTO layer.

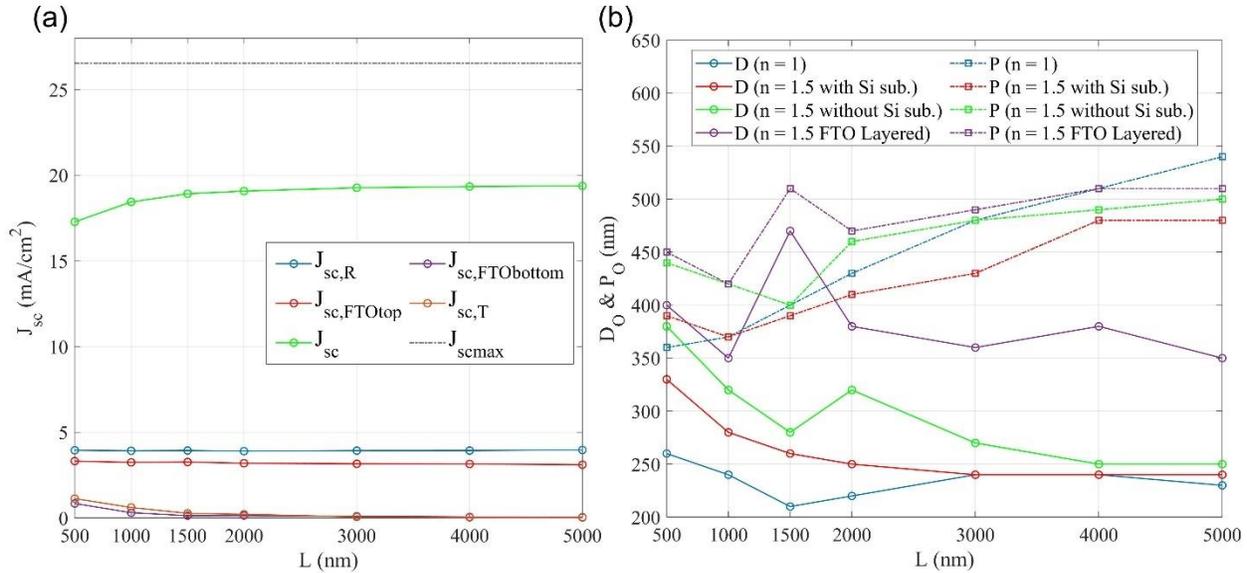

**Figure 8.** a) The $J_{sc}$ calculated for different parts of FTO layered PrvNWs shows that the $J_{sc\_PrvNWs}$ is more influenced by the reflection loss and the absorption of top FTO layer. b) Optimum diameter and period, as a function of PrvNW length, for all the four systems considered in this study (the noticeable jump in both $D_O$ and $P_O$ at $L$ = 1500 nm for the system with FTO layers present originates from minor fringes in the $D$ and $P$ dependent $J_{sc}$, see figure S6 in the Supplementary Information).

For an overall view of the four systems studied, figure 8(b) represents the optimum diameter and period of PrvNWs of FTO layered setup and the previous ones, for each length. Notably, when the FTO layers are present, the optimum diameter is considerably higher, especially at large $L$, where the three other systems converge toward $D_O \approx 250$ nm, while this last system with the FTO layers present shows $D_O \approx 350$ nm at $L = 5000$ nm. As detailed below, this difference in $D_O$ originates from a difference in the diameter dependence of the reflection loss.

For the three first systems (which do not have FTO layers), around $D = D_O \approx 250$ nm, $J_{sc,R}$ is on the order of 0.5 mA/cm² and increases with increasing $D$ (see figure S10 in the Supplementary Information), as expected from the loading of more and more higher refractive index perovskite material into the matrix, subsequently increasing the reflection when $P$ is kept constant. In strong contrast, the $J_{sc,R}$ for the fourth system (which has the FTO layers) is much higher, at around 4 mA/cm2 for the optimum diameter and period, and in strong contrast to the other three systems, it decreases with increasing $D$ (see figure S10 in the Supplementary Information). This decrease in $J_{sc,R}$ with increasing $D$ explains why $D_O$ for this fourth system is pushed to a higher value than for the three other systems. The origin for this diameter behavior for $J_{sc,R}$ must be rooted in the reflection from the top and bottom of the FTO layer (see figure S6 in the Supplementary Information for a schematic). The partial reflection from the top of the layer is not expected to depend strongly on $D$. The decrease in overall reflection loss is understood then as a decrease in the reflection at the bottom interface of the top FTO layer. Indeed, since the $n = 1.5$ material between the PrvNWs has noticeably lower $n$ than the $n \approx 2.18$ of the FTO (see figure S2 in the Supplementary Information), we expect a better refractive index matching to the nanowire matrix if loading heavily with the PrvNWs with $n \approx 2.5$ (see figure S1 in the Supplementary Information). Such heavier loading is achieved with a larger $D$.

To summarize, for the first three systems, the $D_O \approx 250$ nm occurs due to placing the absorption peak from the $HE_{11}$ waveguide mode in the vicinity of the bandgap to enhance the otherwise weak absorption there. However, with the additional interfaces introduced by the top FTO layer, the diameter dependence of the reflection loss overrides the benefit of placing the absorption peak from the $HE_{11}$ mode at the bandgap, instead calling for a noticeably larger optimum $D$ in order to reduce the reflection loss.

## Conclusion

In conclusion, we studied the absorption of light in PrvNW arrays through optics modelling. We found that with optimized diameter for the PrvNWs and period for the array, the PrvNWs can absorb sunlight strongly already at a nanowire length of 1500 nm, with 90% of the maximum short-circuit potential. Importantly, we showed the strong impact that electrical contacting layers can have on the optical response. In the presence of a TCO top contact layer, the optimum diameter shifted to a 100 nm larger value compared to that in a system with a bare PrvNW array without such a layer. The origin of this difference was traced to a reduced reflection from the interface between the TCO layer and the PrvNW array when the PrvNW's diameter is larger. In contrast, for the bare PrvNW arrays, the optimum diameter is set by placing the absorption peak from the $HE_{11}$ waveguide mode to the vicinity of the bandgap wavelength. Thus, our study highlights the importance of optics modelling in the design and optimization of the optical response of PrvNW arrays.

## Data availability statement
All data that support the findings of this study are included within the article (and any supplementary files).

## Conflict of interests
The authors declare no competing interests.

# Supplementary Information
# Optimized Absorption of Light in Perovskite Nanowire Solar Cells

Sina Salimian[1] and Nicklas Anttu[1,*]

[1]Physics, Faculty of Science and Engineering, Åbo Akademi University, FI-20500 Turku, Finland

*E-mail: nicklas.anttu@abo.fi

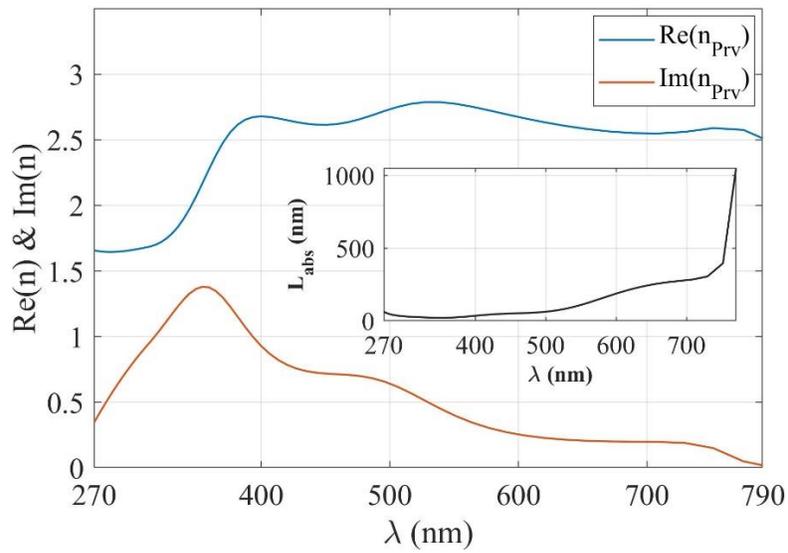

**Figure S1.** The real (blue line, at the top) and the imaginary (red line, at the bottom) part of the refractive index of the perovskite used in our study, as extracted from Ref. [1]. The inset represents the perovskite's absorption length.

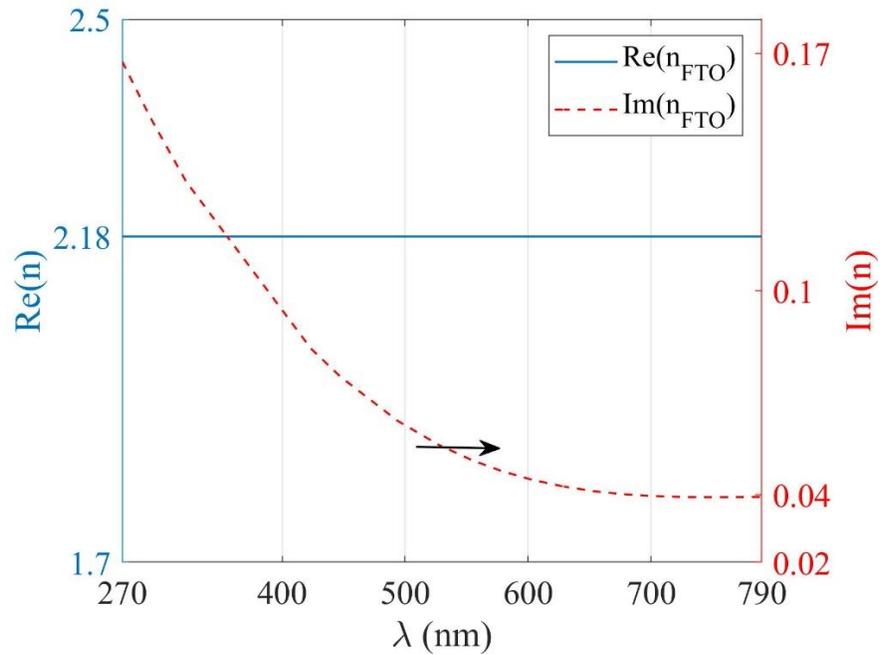

**Figure S2.** The real (blue, solid line) and imaginary (red, dashed line) parts of FTO's refractive index, with values given on the left and the right axis, respectively. The values are extracted from Ref. [2].

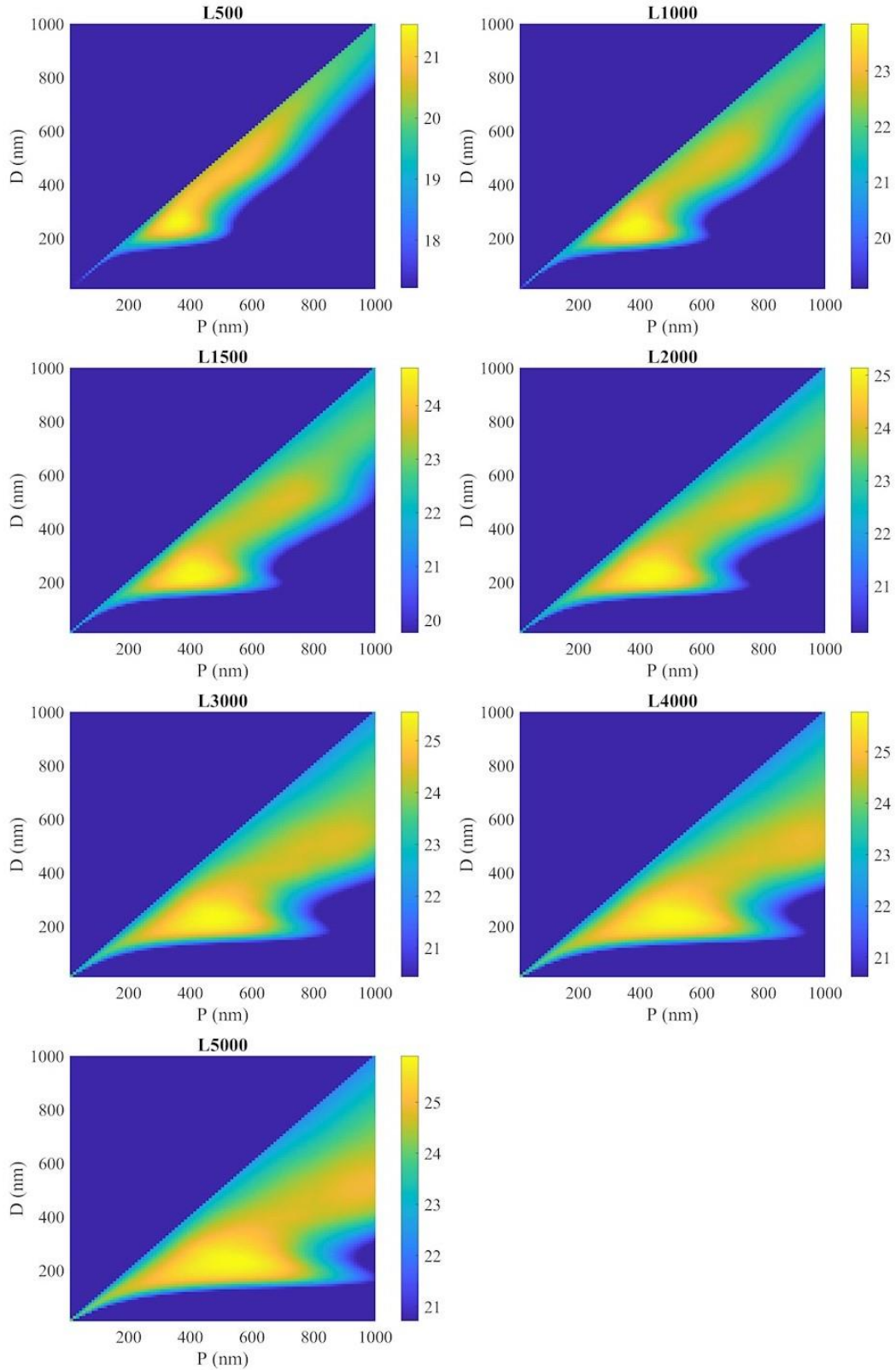

**Figure S3.** $J_{sc}$ in mA/cm$^2$ as a function of $D$ and $P$ for different lengths $L$ (indicated in the title of each subfigure) of the PrvNWs in the first setup (see figure 1(a) for a schematic).

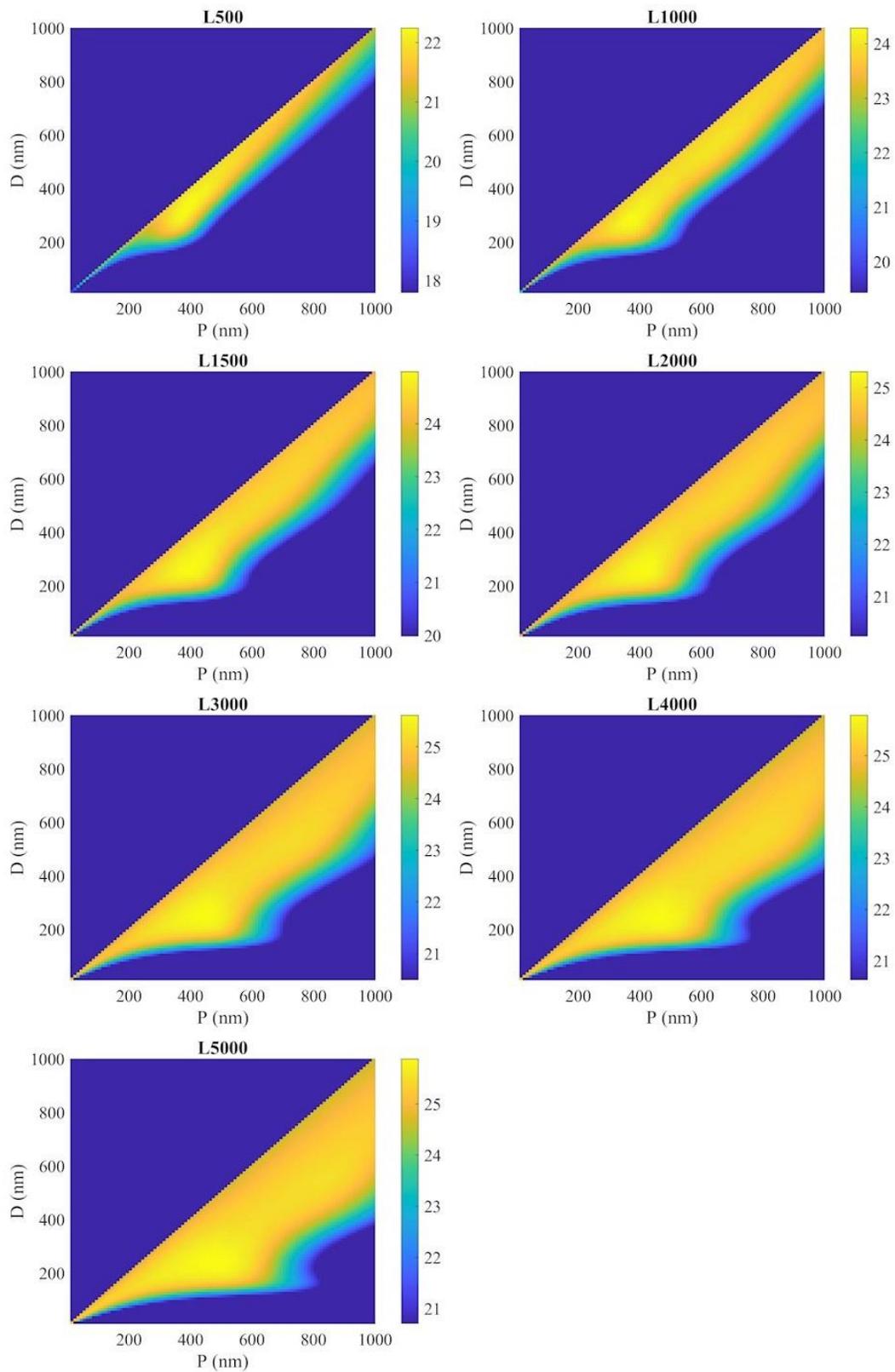

**Figure S4.** $J_{sc}$ in mA/cm$^2$ as a function of $D$ and $P$ for different lengths $L$ (indicated in the title of each subfigure) of PrvNWs in the second setup (see figure 1(b) for a schematic).

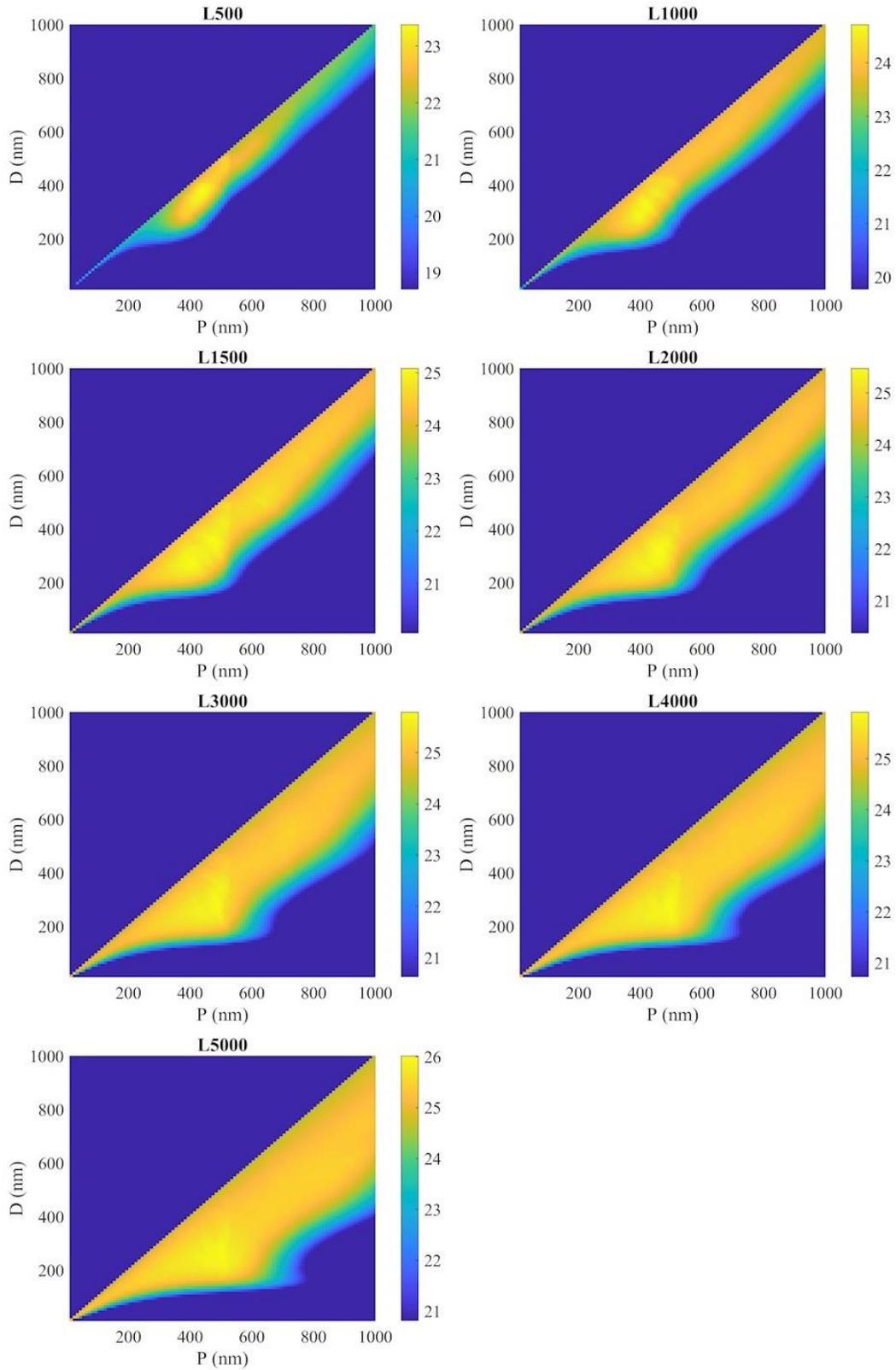

**Figure S5.** $J_{sc}$ in mA/cm$^2$ as a function of $D$ and $P$ for different lengths $L$ (indicated in the title of each subfigure) of PrvNWs in the third setup (see figure 1(c) for a schematic).

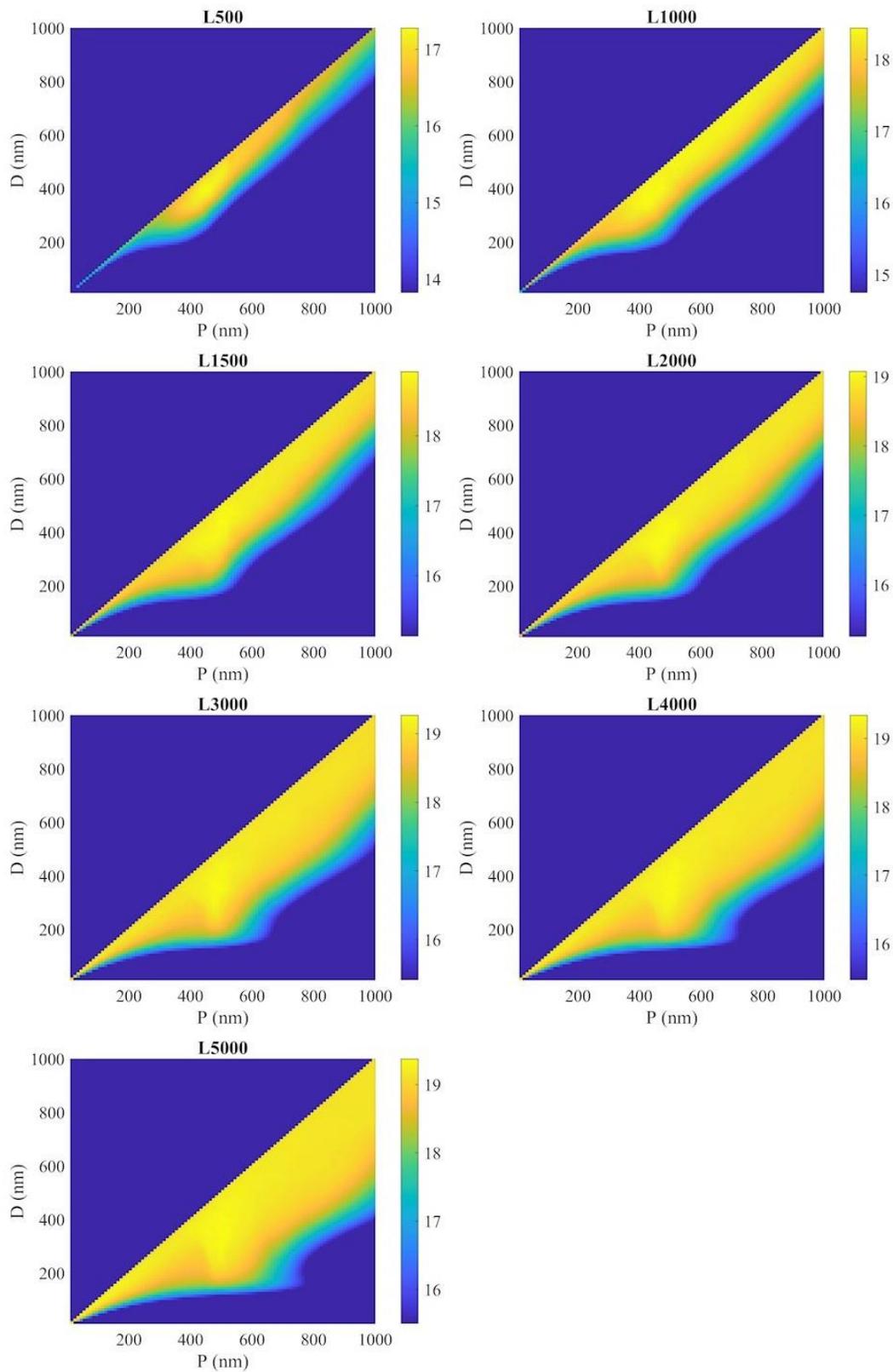

**Figure S6.** $J_{sc}$ in mA/cm$^2$ as a function of $D$ and $P$ for different lengths $L$ (indicated in the title of each subfigure) of PrvNWs in the fourth setup (see figure 1(d) for a schematic).

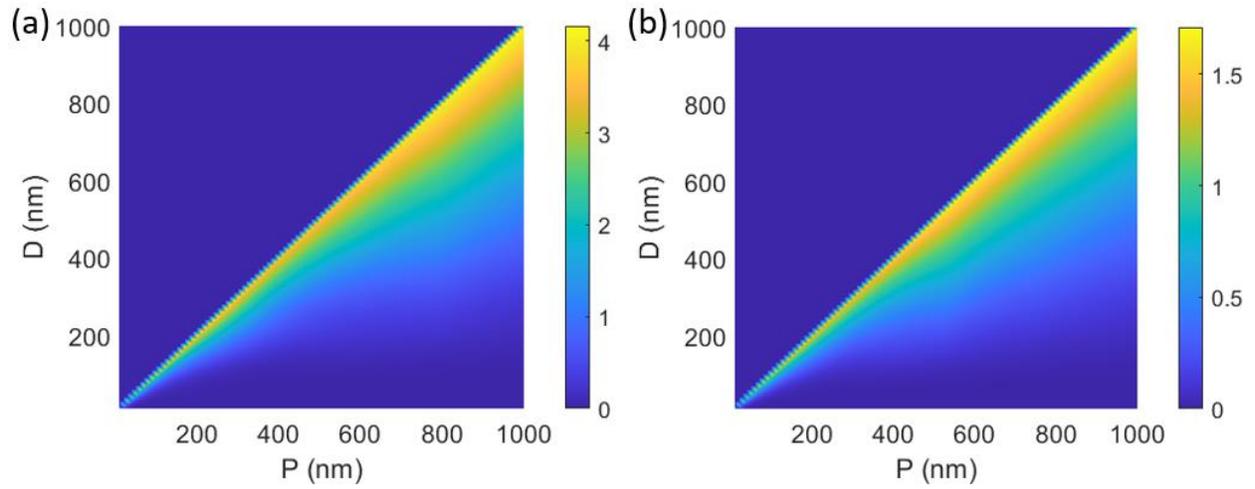

**Figure S7.** The insertion loss in short-circuit current, in mA/cm$^2$, as a function of $D$ and $P$, for (a) the first and (b) the second setup (see figure 1(a)-(b) for schematics, respectively). The results are obtained by placing the bottom PML directly below the PrvNWs, with the PrvNWs continuing into the PML so that no reflection occurs from the bottom side, thus corresponding to $L \to \infty$.

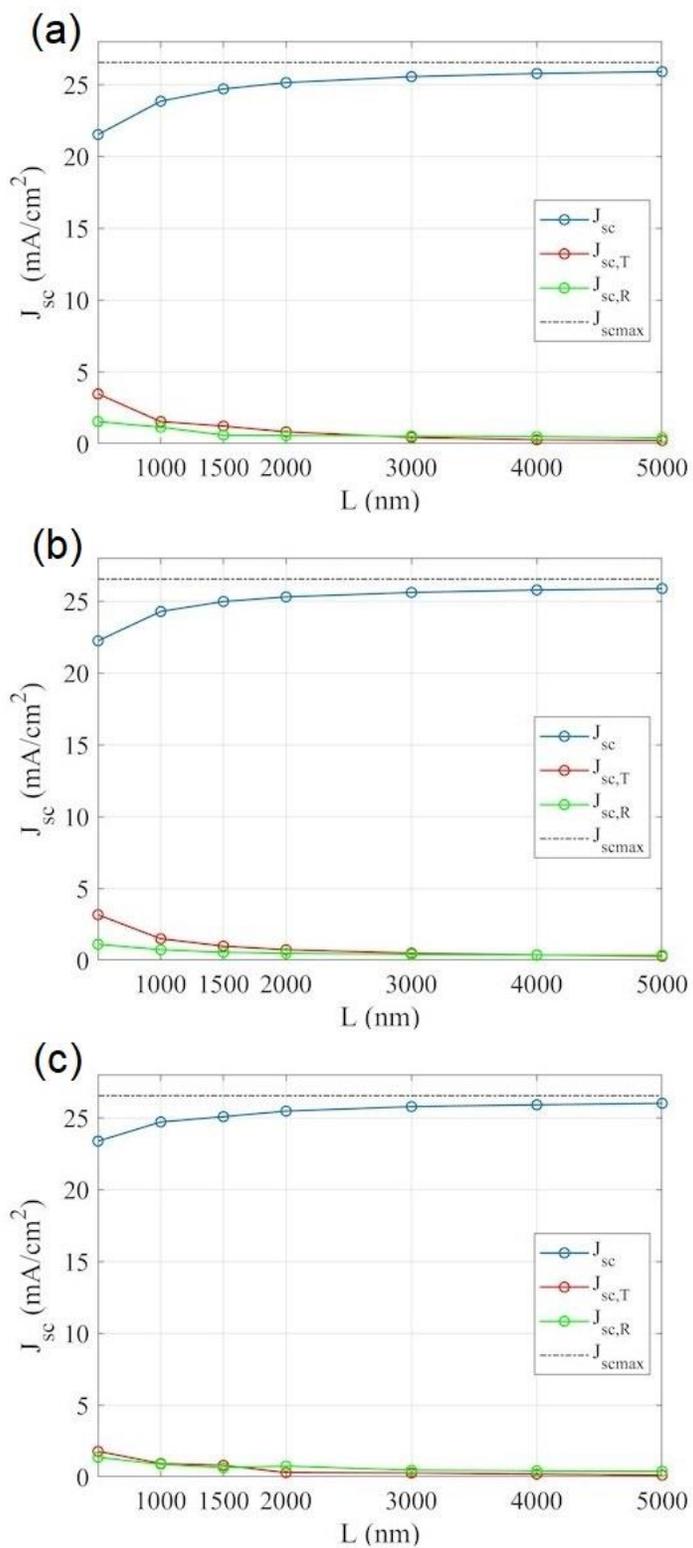

**Figure S8.** Comparison of $J_{sc}$ (absorption in the PrvNWs), $J_{sc,R}$ (reflection loss), and $J_{sc,T}$ (transmission loss) as function of $L$ for optimized $D$ and $P$ for the (a) first, (b) second, and (c) third setup (see figure 1(a)-(c), respectively, in the main text for schematics of the setups).

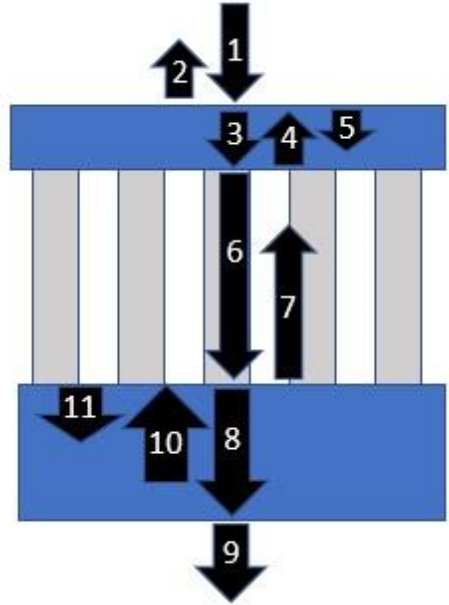

**Figure S9.** Schematic of the scattering of light throughout the fourth setup (see figure 1(d) for a 3D schematic). The incident light (1) impinges on the air/FTO interface resulting in reflection (2) and transmission into the top FTO layer (3). Next, (3) impinges on the FTO/PrvNW-array interface, resulting in reflection (4) and transmission (6) into the PrvNW array (we mark the reflection back into the FTO layer (5), but otherwise terminate marking the continuation of the infinite number of round trips in this FTO layer). (6) results in absorption in the PrvNW array, as well as in reflection (7) and transmission (8) into the bottom FTO layer. (8) gives rise to transmission into the substrate (9), and reflection (10). Note that we here show only the initial steps of the infinitely many round trips within each layer—for each round-trip, the contribution to the reflection and transmission of the system (i.e., $R$ and $T$), as well as to the absorption in the PrvNW, decreases. For example (4) gives rise to transmission back to the top side (not shown in this schematic), which interferes with (2) in $R$. Similarly, (10) gives rise to a reflection back into the bottom layer (11) and transmission into the active layer (not shown in this schematic). It is of importance for our analysis that if (6) is absorbed strongly, then (7) has a negligible contribution to the optical response, in which case the length of the PrvNWs does not have a strong effect on $R$ or the absorption in the top FTO layer.

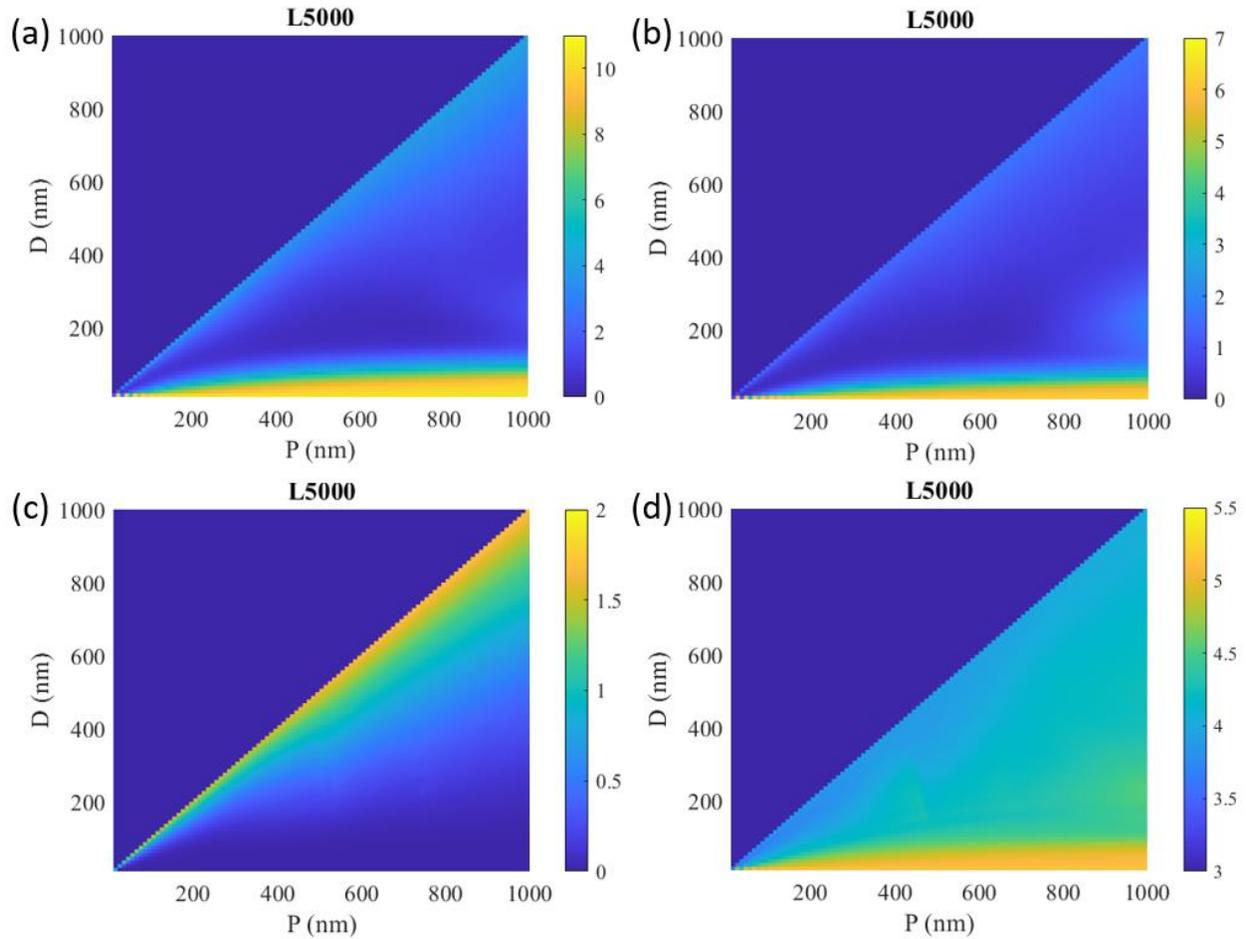

**Figure S10.** $J_{sc,R}$, in mA/cm$^2$, at L = 5000 nm for (a) first, (b) second, (c) third, and (d) fourth setup (see figure 1(a)-(d) for schematics of each setup, respectively). For (a) and (b), the large values for $J_{sc,R}$ at $D \to 0$ originate from the reflection at the interface between the silicon substrate and the $n = 1$ material in (a) and $n = 1.5$ material in (b) [with lower reflection in (b) since the silicon is of higher refractive index and $n = 1.5$ gives a better refractive index matching than $n = 1$]. In (c) we see $J_{sc,R} \to 0$ as $D \to 0$ since then the system is of $n = 1.5$ throughout without any reflection. In (d), when $D \to 0$, the system consists of two FTO layers with $n = 1.5$ material sandwiched between them.